\documentclass[11pt]{article}
\usepackage{jheppub}

\usepackage{amsfonts,amssymb,amsmath,amsthm,url,graphicx,mathtools}
\usepackage{mathrsfs}
\usepackage{bbold}
\usepackage{datetime}
\usepackage{url}
\usepackage{multirow}

\numberwithin{equation}{section}

\DeclareMathOperator{\gl}{\mathfrak{gl}}
\DeclareMathOperator{\sgl}{\mathfrak{sl}}

\DeclareMathOperator{\so}{\mathfrak{so}}
\DeclareMathOperator{\osp}{\mathfrak{osp}}

\DeclareMathOperator{\su}{\mathfrak{su}}
\DeclareMathOperator{\uu}{\mathfrak{u}}

\DeclareMathOperator{\tr}{tr}

\newcommand{\y}{\hat{y}}

\newcommand{\shs}{\mathfrak{shs}}

\newcommand{\be}{\begin{equation}}
\newcommand{\ee}{\end{equation}}

\DeclareMathOperator{\Mat}{Mat}

\numberwithin{equation}{section}

\def\be{\begin{equation}}
\def\ee{\end{equation}}

\title{Extended supersymmetry in AdS$_3$ higher spin theories}

\author{Constantin Candu, Cheng Peng and 
Carl Vollenweider}

\affiliation{
Institut f\"ur Theoretische Physik, ETH Z\"urich \\
CH-8093 Z\"urich, Switzerland}

\emailAdd{canduc@itp.phys.ethz.ch}
\emailAdd{pengch@itp.phys.ethz.ch}
\emailAdd{carlv@itp.phys.ethz.ch}

\abstract{
\noindent We determine the asymptotic symmetry algebra (for fields of low spin) of the $M\times M$ matrix extended Vasiliev theories on AdS$_3$ and find that it agrees with the $\mathcal{W}$-algebra of their
proposed coset duals. Previously it was noticed that for $M=2$ the supersymmetry increases from $\mathcal{N}=2$ to $\mathcal{N}=4$. We study more systematically this type of supersymmetry enhancements and find that, although the higher spin algebra has extended supersymmetry for all $M\geq 2$, the corresponding asymptotic symmetry algebra fails to be superconformal except for $M=2$, when it has 
large $\mathcal{N}=4$ superconformal symmetry.
Moreover, we find that the Vasiliev theories based on $\shs^E\! \left( \mathcal{N} \vert 2, \mathbb{R} \right)$ are special cases of
the matrix extended higher spin theories, and hence have the same supersymmetry properties.
}

\date{\today}

\allowdisplaybreaks

\begin{document}
\maketitle


\section{Introduction} \label{sec:intro}

Theories with higher spin symmetry have long been speculated to be related to the tensionless limit of string theory\cite{Gross:1988ue, Witten:1988zd, Moore:1993qe}.
Recently, concrete proposals have been made for embeddings of higher spin/vector model dualities into string/gauge dualities. 
In \cite{Chang:2012kt} a particular incarnation of an AdS$_4$/Chern-Simons vector model duality \cite{Aharony:2011jz, Giombi:2011kc} has been put forward and linked to 
string theory \cite{Aharony:2008gk, Aharony:2008ug}.
The duality of \cite{Chang:2012kt} relates Vasiliev higher spin theories extended with $\mathrm{U}(M)$ Chan-Paton factors and subject to $\mathcal{N}=6$ boundary conditions to the large $N$ limit of
$\mathrm{U}(M) \times \mathrm{U}(N)$ ABJ theories. The $\mathrm{O}(N)$ vector model duality of Klebanov \& Polyakov \cite{Klebanov:2002ja} (see also \cite{Giombi:2013fka} for recent progress on this topic) is recovered for $M=1$, the relation to string theory
is established in the large $M$ limit.

In one dimension lower,  
there have been successful attempts to obtain similar embeddings in the context of AdS$_3$/minimal model holography \cite{Gaberdiel:2010pz} (see  \cite{Gaberdiel:2012uj} for a review).
In \cite{Gaberdiel:2013vva, Creutzig:2013tja} dualities between the matrix extended Vasiliev  higher spin theories  on AdS$_3$  based on the higher spin algebra $\shs_M[\mu]$ \cite{Prokushkin:1998bq, Prokushkin:1998vn}
and the large $N$ limit of  $\mathrm{SU}(N+M)/ \mathrm{SU}(N) \times \mathrm{U}(1)$ type cosets of supersymmetric Wess-Zumino-Witten models have been proposed. A pivotal observation has been the finding in \cite{Gaberdiel:2013vva} 
that for the special value $M=2$ both theories exhibit enhanced large $\mathcal{N}=4$ superconformal symmetry, see also \cite{Beccaria:2014jra, Gaberdiel:2014yla, Ahn:2013oya, Ahn:2014via}. 
Then in \cite{Gaberdiel:2014cha}, Gaberdiel \& Gopakumar have convincingly demonstrated  that these cosets with $M=2$  can be embedded into a
free symmetric orbifold CFT with small $\mathcal{N}=4$ supersymmetry,
which is believed to be dual to the tensionless limit of string theory on AdS$_3 \times S^3 \times \mathbb{T}^4$ \cite{David:2002wn}. They have proven that in the large level limit, in which the large $\mathcal{N}=4$ superconformal algebra degenerates to the small $\mathcal{N}=4$ superconformal algebra,  the 
(perturbative part of the) coset
can be described as a closed subsector of the (untwisted sector of the) free symmetric orbifold (see also \cite{Gaberdiel:2014vca} for preliminary work).
In particular, this implies that the  above higher spin theories with a  Chan-Paton factor of size $M=2$ can be viewed as a subsector of 
the tensionless limit of string theory on AdS$_3 \times S^3 \times \mathbb{T}^4$.

Finding a similar string theory interpretation for the  higher spin/coset theories with $M>2$ turned out to be very difficult because not only do
these theories  lack the $\mathcal{N}=4$ superconformal symmetry \cite{Candu:2013fta}, but in fact 
they do not appear to be supersymmetric at all.
The situation improved in  \cite{Creutzig:2014ula}, where it was noticed that the Kazama-Suzuki type variant of the above cosets, which is in general $\mathcal{N}=2$, 
admits at a special value of the level $k$, but still for general values of $M$ and $N$,  extended $\mathcal{N}=3$ superconformal symmetry.
As emphasized in \cite{Creutzig:2014ula}, this important observation leaves only a few candidate string duals. 
This paves the way for a more detailed study in which convincing evidence for (or against) the duality with string theory can be produced.

In this work we shall study and test the matrix extended duality of \cite{Gaberdiel:2013vva, Creutzig:2013tja} for general  $M$.
At the level of partition functions this duality has already been checked in \cite{Creutzig:2013tja, Candu:2013fta}, where it has been shown
that the 1-loop partition function of the extended Vasiliev theory matches precisely the coset partition function in the 't~Hooft limit (with null states removed).
In this paper,  we shall compute the asymptotic symmetry algebra at low spins (for $M=4$), and check the agreement with the predictions from the dual coset theory.
Our computations reproduce similar results  obtained by different methods in \cite{Creutzig:2013tja, Creutzig:2014ula}.

The main objective of this paper is to study in a more systematic way the supersymmetry of the Vasiliev theory based on the higher spin algebra $\shs_M[\mu]$.
As it turns out, one can find supersymmetry subalgebras of $\shs_M[\mu]$ for any $M>2$,\footnote{The $\mathcal{N}=4$ supersymmetry construction of \cite{Gaberdiel:2013vva}
generalizes to $\shs_M[\mu]$ because of the
natural embedding  $\shs_2[\mu]\subset \shs_M[\mu]$ for $M>2$.}
but, quite surprisingly (see e.g.~\cite{Henneaux:2012ny}),
this does not automatically guarantee that the algebra of asymptotic symmetries is superconformal.
In fact, our analysis suggests that this is only possible for $M=1$ and $M=2$, when the asymptotic symmetry algebra has $\mathcal{N}=2$ and
large $\mathcal{N}=4$ superconformal symmetry.\footnote{This does not contradict the coset construction \cite{Creutzig:2014ula} of $\mathcal{N}=3$ supersymmetry for any $M$ because on the higher spin side this corresponds to an extension of the asymptotic symmetry algebra of $\mathfrak{shs}_M[\tfrac{1}{2}]$.} This has been predicted from the coset point of view in \cite{Candu:2013fta} --- although without providing any details of the computations.

Another goal of this work is to study the relation between the above matrix extended Vasiliev theories and the higher spin theories based on the algebra $\shs^E\! \left( \mathcal{N} \vert 2, \mathbb{R} \right)$
\cite{Vasiliev:1986qx,Blencowe:1988gj}.
These theories have recently been investigated in \cite{Henneaux:2012ny} and \cite{Creutzig:2014ula}.
In \cite{Henneaux:2012ny} Henneaux et al. determined its asymptotic symmetry algebra and mentioned possible relations to string theory; and in \cite{Creutzig:2014ula} (generalizing the earlier
work of \cite{Beccaria:2013wqa}) several dualities 
involving these theories for \emph{odd} $\mathcal{N}$ have been proposed.
In the present paper, we shall propose coset duals also for \emph{even} $\mathcal{N}$, and this will also allow us to shed some light on the relation between the different coset duals proposed in \cite{Creutzig:2014ula} 
in the case of odd $\mathcal{N}$. 

The paper is organized as follows. In section~\ref{sec:mat} we introduce the matrix extended higher spin algebras 
on which the Vasiliev theories in the duality of \cite{Gaberdiel:2013vva, Creutzig:2013tja} are based, and show that they are related 
to the $\shs^E\! \left( \mathcal{N} \vert 2, \mathbb{R} \right)$ higher spin algebras
and contain the Lie superalgebra  $\mathfrak{osp}( \mathcal{N} |2)$ as a subalgebra.
In section~\ref{sec:check} we do the asymptotic symmetry analysis and compare it to the corresponding coset computation.
The obstructions to the existence of supersymmetry are studied more systematically in section~\ref{sec:susy} and finally, section~\ref{sec:concl} contains a short summary of our results and conclusions.
The OPEs used to perform the coset computation can be found in appendix \ref{sec:cosetopes}.


\section{The matrix extended higher spin algebras}\label{sec:mat}

In this section we shall introduce the higher spin algebras of the matrix extended Vasiliev theories 
\cite{Prokushkin:1998bq, Prokushkin:1998vn}. In addition, we will show that these are related to the  
$\shs^E\! \left( \mathcal{N} \vert 2, \mathbb{R} \right)$
higher spin algebras \cite{Vasiliev:1986qx, Blencowe:1988gj} and contain the Lie superalgebra  $\mathfrak{osp}( \mathcal{N} |2)$ as a subalgebra.

\subsection{Construction}\label{sec:constr}

Let us begin by defining the $\mathcal{N} = 2$ higher spin algebra $\shs[\mu]$ \cite{Bergshoeff:1991dz}. It is well-known that $\shs[\mu]$ can be obtained from the associative algebra
\begin{equation}
sB[\mu] = \frac{U(\osp(1|2))}{\langle \mathrm{Cas}-\tfrac{1}{4}\mu(\mu-1)\mathbb{1}\rangle} \,, 
\label{eq:sB}
\end{equation}
where $U(\osp(1|2))$ is the universal enveloping algebra and $\mathrm{Cas}$ is the Casimir of the Lie superalgebra $\osp(1|2)$.
A convenient basis for $sB[\mu]$ \cite{Prokushkin:1998bq, Prokushkin:1998vn}  can be realized in terms of the oscillators $\y_1$, $\y_2$ and $k$ as
\begin{equation} \label{basissB}
V_m^{(s)\pm} \propto \y_{(\alpha_1...}\, \y_{\alpha_{l})}P_\pm\ ,
\end{equation}
where $s=\frac{l}{2}+1$ , $l \in \mathbb{N}$ and  $2m=\#\y_1-\#\y_2$ takes values in the range $-s+1\leq m \leq s-1$, 
while $P_\pm: =(\mathbb{1} \pm k)/2$ are projectors due to the following properties of the operator $k$
\be\label{krel}
k \y_\alpha = - \y_\alpha k\ ,\quad k^2 = 1\,.
\ee
These relations are supplemented by those for the oscillators $\y_{\alpha}$
\begin{equation}
[\y_\alpha,\y_\beta]=\y_\alpha\y_\beta-\y_\beta\y_\alpha=2i\epsilon_{\alpha\beta}(\mathbb{1}+\nu k) \ , 
\label{eq:oscalg}
\end{equation}
where $\epsilon_{\alpha\beta}= - \epsilon_{\beta\alpha}$, $\epsilon_{12}=1$, and $\nu=2\mu-1$.

A natural grading $|\cdot|$ on $sB[\mu]$ is induced by the operator $k$ due to the relations~\eqref{krel}. Indeed, the generators $V^{(s)\pm}_m$ with integer (half odd-integer) index $s$ are even (odd) with respect to $k$. In this paper we adopt standard convention and call them bosonic  and fermionic, respectively.  
Then we can define the following graded Lie bracket on $sB[\mu]$
\be\label{supalg}
[a,b\} := ab - (-1)^{|a||b|}ba\, .
\ee

Notice that there is a distinguished $\sgl(2)$ subalgebra of $sB[\mu]$, namely the one corresponding to the 
bosonic $\sgl(2)$ subalgebra of the original $\osp(1|2)$ from \eqref{eq:sB}. It is spanned by
\begin{equation}
L_1 = \tfrac{1}{4i}\y_1^2\ ,\quad L_{-1} = \tfrac{1}{4i}\y_2^2\ , \quad L_0 =\tfrac{1}{8i}(\y_1\y_2+\y_2\y_1)\,,
\label{eq:emt}
\end{equation}
which correspond to the $s=2$ piece of \eqref{basissB}. With respect to this $\sgl(2)$ subalgebra the $V_m^{(s)\pm}$ satisfy
the commutation relations
\begin{equation}
[L_m,V^{(s)\pm}_n] = [m(s-1)-n]V^{(s)\pm}_{m+n} \ ,
\end{equation}
and for this reason the index $s$ in \eqref{basissB} is usually referred to as the `spin' of the generator.

The Lie superalgebra $sB[\mu]$ decomposes (as a vector space)
into the two subspaces
\begin{equation}
sB[\mu]  = 
\mathbb{C}\mathbb{1}\oplus\mathrm{span}_{s>1}(J_0,V_m^{(s)\pm})
\label{eq:sBintoosp}
\end{equation}
with $J_0 := -\tfrac{1}{2}(\nu 
+k)$,
and one can check that
the first subspace $\mathbb{C} \mathbb{1}$ in \eqref{eq:sBintoosp} is not generated by the super-commutators on the second subspace. Hence, the subspace $\mathrm{span}_{s>1}(J_0,V_m^{(s)\pm})$  endowed with the graded Lie bracket \eqref{supalg} constitutes a Lie subsuperalgebra --- the Lie superalgebra,
which we denote by $\shs[\mu]$. Observe that $\shs[\mu]$
contains an $\mathcal N=2$  supersymmetry subalgebra that is generated by $J_0$, $L_0$, $L_{\pm 1}$ and the four spin  $s= 3/2$ generators
\begin{equation}
G^\pm_{1/2}=\tfrac{1}{2\sqrt{2}}e^{-i\pi/4}\y_1 (1\pm k)\ ,\quad G^{\pm}_{-1/2}=\tfrac{1}{2\sqrt{2}}e^{-i\pi/4}\y_2 (1\pm k)\ .
\label{eq:n2supercharges}
\end{equation}

We shall now extend the algebras $sB[\mu]$ and $\shs[\mu]$ by tensoring them with the algebra $\Mat_M$ of complex $M\times M$ matrices.
First, let us define
\begin{equation}
sB[\mu]_M := sB[\mu]\otimes \Mat_M\ .
\label{eq:extsB}
\end{equation}
Clearly, $sB_M[\mu]$ is an associative superalgebra, which can be turned into a Lie superalgebra in the usual way.
It inherits a natural $\sgl(2)$ subalgebra
$\mathbb{L}_m \equiv L_ m  \otimes \mathbb{1}_M$,
a grading $|a\otimes A|= |a|$ and a 
trace
$\tr a\otimes A = \tr a \tr A$, where the trace on $sB[\mu]$ is defined as the projection on the identity from \eqref{eq:sBintoosp}.
A convenient basis for $sB_M[\mu]$ is given by
\be\label{matrixbasis}
V^{(s)ij\pm}_m \propto  V^{(s)\pm}_m \otimes E_{ij}\,,
\ee
where the $(E_{ij})_{mn}=\delta_{im}\delta_{jn}$ are the standard matrix basis elements. 
Eliminating the $\mathfrak{gl}(1)$ subalgebra spanned by $\mathbb{1}\otimes\mathbb{1}_M$,  we get the so-called matrix extension of $\shs[\mu]$
\begin{equation}
\shs_M[\mu] := \mathbb{1}\otimes \mathfrak{sl}(M)\; \oplus\; \shs[\mu]\otimes \mathbb{1}_M\;\oplus\;\shs[\mu]\otimes \mathfrak{sl}(M)
\label{eq:extshs}
\end{equation}
Its  spin $s=1$ subspace is a direct sum of the two  mutually commuting subalgebras
\begin{equation}
\mathfrak{sl}(M)_\pm:=P_\pm\otimes \mathfrak{sl}(M)
\label{eq:slnpm}
\end{equation}
together with the $\gl(1)$ subalgebra generated by $J_0\otimes \mathbb{1}_M = -\tfrac{1}{2}(\nu 
+k) \otimes \mathbb{1}_M$. Using the explicit basis \eqref{basissB}, it can be checked
that the spin $s$ subspaces of $\mathfrak{shs}_M[\mu]$ decompose into the
following multiplets of $\mathfrak{sl}(M)_+\oplus \mathfrak{sl}(M)_-\oplus \gl(1)$
\begin{align}\label{eq:dec_shsn}
s=1:&\quad  (adj, 0)_0 \oplus(0,adj)_0\oplus (0,0)_0\\ \notag
s\in \mathbb{N}+\tfrac{1}{2}:&\quad (f,f^*)_{-1}\oplus(f^*,f)_{1} \\ \notag
s\in\mathbb{N}+1:&\quad (adj,0)_0\oplus(0,adj)_0\oplus 2 (0,0)_0\ ,
\end{align}
where $f$ and  $f^*$ is the fundamental and anti-fundamental representation of $\mathfrak{sl}(M)$, respectively, $adj$ is the adjoint representation and the lower index denotes the $J_0$ charge.

\subsection{Relation to the \texorpdfstring{$\shs^E\! \left( \mathcal{N} \vert 2, \mathbb{R} \right)$}{shsE(N,2,R)} higher spin algebras } \label{sec:hsrel}

Our next aim is to show that the higher spin algebras $\shs_M[\mu]$ at the special value $\mu = 1/2$
are closely related to another class of higher spin algebras, usually denoted 
by $\shs^E\! \left( \mathcal{N} \vert 2, \mathbb{R} \right)$, which were introduced by Vasiliev in 
\cite{Vasiliev:1986qx} (see also \cite{Blencowe:1988gj}). We shall see that the following isomorphisms hold
\begin{align}
\label{eq:isoeven}
\shs_M[\tfrac{1}{2}]  &\cong \shs^E\! \left( \mathcal{N} \vert 2, \mathbb{R} \right) \quad 
\mathrm{with} \ M=2^{\mathcal{N} / 2 -1} \ \mathrm{and} \ \mathcal{N} \ \mathrm{even} \ , \\
\left( \shs_M[\tfrac{1}{2}] \right)^{\mathbb{Z}_2} &\cong \shs^E\! \left( \mathcal{N} \vert 2, \mathbb{R} \right)   \quad 
\mathrm{with} \ M=2^{(\mathcal{N} -1) / 2}  \ \mathrm{and} \ \mathcal{N} \ \mathrm{odd} \ , \label{eq:isoodd}
\end{align}
where $\left( \shs_M[\tfrac{1}{2}] \right)^{\mathbb{Z}_2} \subset \shs_M[\tfrac{1}{2}]$ is the subalgebra fixed by
the $\mathbb{Z}_2 $-automorphism $k \to -k$. 
As explained in the Introduction
these isomorphisms will eventually allow us to relate the results of our paper to the
recent works of \cite{Henneaux:2012ny} and \cite{Creutzig:2014ula}, cf.\ section~\ref{sec:rel}.

The higher spin algebra $\shs^E\! \left( \mathcal{N} \vert 2, \mathbb{R} \right)$ can be realized, similarly as $\shs_M[\mu]$, in terms of oscillators.
To this end, let us introduce two bosonic oscillators $\y_\alpha$, $\alpha=1,2$ 
as well as $\mathcal{N}$ fermionic oscillators $\hat{\psi}_i$, $i=1,\ldots, \mathcal{N}$, which satisfy the
relations\footnote{For the sake of consistency, we chose to employ the same conventions as 
in section~\ref{sec:constr} to describe the oscillators.
Note, however, that there exists an alternative description of $\shs^E\! \left( \mathcal{N} \vert 2, \mathbb{R} \right)$ (see e.\,g.\ \cite{Vasiliev:1986qx} or \cite{Henneaux:2012ny})
in terms of symbols and a star product defined as
\[
(f\star g)(z)\equiv f(z') \exp(i\epsilon_{\alpha\beta}\frac{\overleftarrow{\partial}}{\partial y'_\alpha} \frac{\overrightarrow{\partial}}{\partial y''_\beta}+\delta_{ij}\frac{\overleftarrow{\partial}}{\partial \psi'_i} \frac{\overrightarrow{\partial}}{\partial \psi''_j} ) g(z'')\bigg|_{z',z''\to z}\,,
\]
where we have used the shorthanded notation $z= (y_1,y_2,\psi_1,\ldots,\psi_\mathcal{N})$.}
\begin{equation}
\left[ \y_\alpha , \y_\beta \right] = 2i\epsilon_{\alpha\beta}\ ,\qquad  \{ \hat{\psi}_i , \hat{\psi}_j \}  = 2 \delta_{ij}\ ,\qquad [ \y_\alpha , \hat{\psi}_i ] = 0 \ .
\end{equation}
Then $\shs^E\! \left( \mathcal{N} \vert 2, \mathbb{R} \right)$ can be defined as the oscillator algebra generated by all 
 polynomials in these oscillators subjected to the constraint 
\begin{equation} \label{eq:chooseeven}
 p+q+K \in 2 \mathbb{Z} \ ,
\end{equation}
where we denoted by $p$ and  $q$ the power of the oscillators $\y_1$ and $\y_2$, respectively, while $K$ counts the total number of 
fermionic oscillators $\hat{\psi}_i$.

The higher spin algebra $\shs^E\! \left( \mathcal{N} \vert 2, \mathbb{R} \right)$ defined as above 
can be thought of as a `polynomial' extension of the Lie superalgebra $\osp( \mathcal{N} \vert 2)$. Indeed, it is easy to check that 
the 
polynomials of degree two in the oscillators
\begin{equation}X_{ij}\equiv \frac{1}{2} \hat{\psi}_i \hat{\psi}_j \ ,  \qquad X_{\alpha j}\equiv \frac{1}{2} \y_\alpha \hat{\psi}_i \ , \qquad 
Y_{\alpha\beta}\equiv -\frac{i}{2} \y_\alpha \y_\beta
\end{equation}
form an $\osp( \mathcal{N} \vert 2)$ subalgebra.

Our next goal is to prove the above isomorphisms \eqref{eq:isoeven} and \eqref{eq:isoodd}. 
In a first step, we shall show that the spectra of the algebras agree. In fact, it is a simple exercise to count the number 
of generators of the higher spin algebra 
$\shs^E\! \left( \mathcal{N} \vert 2, \mathbb{R} \right)$: From the fact that the `conformal'
$\sgl(2)$ subalgebra is spanned by the generators $Y_{\alpha\beta}$, it follows that the (modes of the) generators of spin $s$  are given by the 
products of $p$ $\y_{1}$-oscillators and $q$ $\y_{2}$-oscillators such that $p+q=2(s-1)$,
where the $2s- 1$   pairs $(p,q)$ correspond to the $2s- 1$ different modes of the spin-$s$ generators.
Furthermore, there are $2^{\mathcal{N}-1}$ allowed ``configurations" of the fermionic oscillators at each fixed spin. (The exception being the generators of spin 1
from which we exclude the identity.) That is the higher spin algebra consists of 
$2^{\mathcal{N} -1 }$ generators of spin $s \geq 3/2$ (and $2^{\mathcal{N} -1 } -1 $ generators of spin 1).

We can compare this now with the spectra of the matrix extended higher spin algebras. 
It follows e.\,g.\ from eq.~\eqref{eq:dec_shsn} that the higher spin algebra $\shs_M[\mu]$ contains 
$2 M^2$ generators of spin $s \geq 3/2$ (and $2 M^2 -1$ generators of spin 1). Hence, due to the relation
$M=2^{\mathcal{N} / 2 -1}$, 
we conclude that indeed the spectra of generators of the two algebras in \eqref{eq:isoeven} agree, and similarly in \eqref{eq:isoodd}.

To complete the proof of the above isomorphisms, recall that the irreducible representations of the Clifford algebra with $\mathcal{N}$ generators
are of  dimension $M=2^{\lfloor \mathcal{N}/2 \rfloor}$.
In particular, this means that the matrix part of $\left( \shs_M[\tfrac{1}{2}] \right)^{\mathbb{Z}_2}$ 
with $M = 2^{\ell}$ can accommodate  at most $\mathcal{N}=2\ell+1$ Gamma matrices, which we denote by $\Gamma^1,\dots,\Gamma^{2\ell+1}$. They satisfy the Clifford algebra relations
\begin{equation} \label{eq:cliff}
\{ \Gamma^p , \Gamma^q \} = 2 \delta_{pq} \mathbb{1}_M\ .
\end{equation}

The isomorphism \eqref{eq:isoodd} with $\mathcal{N}=2\ell+1$ can be constructed as follows.
First, notice that $\shs^E\! \left( \mathcal{N} \vert 2, \mathbb{R} \right)$ is generated by the elements $\y_\alpha\y_\beta$, $\y_\alpha  \hat{\psi}_{p}$ and $\hat{\psi}_{p} \hat{\psi}_{q}$.
Then one can easily check that the following mapping of these generators into elements of $\mathfrak{shs}_M[\tfrac{1}{2}]$
\begin{equation} \label{eq:psigamma}
 \y_\alpha\y_\beta  \mapsto  \y_\alpha\y_\beta\otimes \mathbb{1}_M\ ,\qquad
\y_\alpha  \hat{\psi}_{p} \mapsto \y_\alpha  \otimes \Gamma^{p} \ ,\qquad
\hat{\psi}_{p} \hat{\psi}_{q} \mapsto \mathbb{1}  \otimes \Gamma^{p}\Gamma^{q}
\end{equation}
is compatible with the associative product. Since the two algebras have the same spectrum at every spin $s$, it follows that this algebra homomorphism is also an isomorphism.

The isomorphism \eqref{eq:isoeven} for $\mathcal{N}=2\ell+2$ must be constructed a bit differently because one has one more $\hat{\psi}$ than Gamma matrices.
Explicitly, in addition to eqs.~\eqref{eq:psigamma} with $p,q\leq 2\ell+1$ one must add the following identifications
\begin{equation} \label{eq:npsigamma}
\y_\alpha \hat{\psi}_{\mathcal{N}}  \mapsto \y_\alpha k \otimes \mathbb{1}_M \ ,\qquad
\hat{\psi}_{p} \hat{\psi}_{\mathcal{N}} \mapsto k  \otimes \Gamma^{p}\ .
\end{equation}
Once again, it is straightforward to check the compatibility of this mapping with the associative product.

\subsection{Relation of the \texorpdfstring{$\mathcal{W}$}{W}-algebras}

\label{sec:rel}

In the previous section we have seen that for even $\mathcal{N}$ the higher spin algebras $\shs^E\! \left( \mathcal{N} \vert 2, \mathbb{R} \right)$ are isomorphic to 
the higher spin algebras $\shs_M[\mu]$ for $\mu = 1/2$ and $M$ given as in \eqref{eq:isoeven}. This means
that all the results, which we will obtain in the following sections for the matrix extended higher spin theories, also hold for the higher spin theories based on $\shs^E\! \left( \mathcal{N} \vert 2, \mathbb{R} \right)$.
In particular, the isomorphism~\eqref{eq:isoeven} together with the  holographic duality of~\cite{Gaberdiel:2013vva, Creutzig:2013tja} implies that the coset duals of the $\shs^E\! \left( \mathcal{N} \vert 2, \mathbb{R} \right)$
higher spin theories with $\mathcal{N}=2\ell+2$ are given by the large $N$ limit of the cosets \eqref{eq:lcoset} with $k=N$ and $M=2^\ell$.
This resolves one of the questions raised in \cite{Henneaux:2012ny}.
Another immediate consequence from the isomorphism~\eqref{eq:isoeven} is that the $\lambda$-deformed $\shs^E\! \left( \mathcal{N} \vert 2, \mathbb{R} \right)$ higher spin theories
are simply given by the higher spin theories based on $\shs_M[\lambda]$ (cf.\ the discussion in \cite{Henneaux:2012ny} and also \cite{Gomez:2014dwa}). 

On the other hand, the case of odd $\mathcal{N}$ is more subtle. This is because the isomorphism~\eqref{eq:isoodd} between higher spin algebras is not guaranteed to lift to an isomorphism
between asymptotic symmetry algebras (ASA), i.e.\ the isomorphism
\begin{equation}\label{eq:isom?}
\mathrm{ASA} \left( \shs_M[\tfrac{1}{2}] \right)^{\mathbb{Z}_2}  
\stackrel{?}{\cong} \left( \mathrm{ASA}\  \shs_M[\tfrac{1}{2}] \right)^{\mathbb{Z}_2}
\end{equation}
is far from obvious. An illustrative example of a higher spin algebra embedding which does not promote to an ASA embedding is discussed in section~\ref{sec:asym}. 
Let us remark that one of  the proposals in \cite{Creutzig:2014ula}  that the coset duals of the higher spin theories based on $\shs^E\! \left( \mathcal{N} \vert 2, \mathbb{R} \right)$ for $\mathcal{N}$ odd are
given by a $\mathbb{Z}_2$-orbifold of the cosets \eqref{eq:lcoset} relies on the validity of the above isomorphism.
Although our asymptotic symmetry analysis in section \ref{sec:asym} agrees with this conjecture, we think that a more thorough study is required to clarify the relation between the asymptotic symmetry algebra and
the $\mathbb{Z}_2$-orbifold.
This seems appropriate especially because a different proposal in \cite{Creutzig:2014ula} for the coset duals requires that the l.h.s.\ of eq.~\eqref{eq:isom?} be also an extension of the asymptotic symmetry algebra of $\mathfrak{hs}_M[\tfrac{1}{2}]$.


%

\subsection{Supersymmetry subalgebras} \label{sec:subalg}

In this section we shall construct explicitly $\mathfrak{osp}( \mathcal{N} |2)$ subalgebras 
of $\shs_M[\mu]$ with $M=2^\ell$ and $\mu = 1/2$ for $\mathcal{N}=2\ell+1$ and $\mathcal{N}=2\ell+2$.
Let us emphasize that setting $\mu=1/2$ is indeed
 necessary for the reasons that we will explain in a moment.

We consider the case of $\mathcal{N}=2\ell+1$ first. Let us define the spin $s=1$ generators 
\begin{equation}
J^{(kl)}_0 = \mathbb{1}\otimes \frac{[\Gamma^k,\Gamma^l]}{4}\\
\label{eq:hsJ}
\end{equation}
and the spin $s=3/2$ generators
\begin{equation}\label{eq:ansatz_G}
G^p_r = \frac{e^{-i\pi/4}}{2}\, \y_\alpha \otimes \Gamma^p\ .
\end{equation}
Here  $1 \leq k,l,p\leq \mathcal{N}$ 
and $\Gamma^p$ are the maximum number of $2\ell+1$  Gamma matrices, which can be realized in $M=2^{\ell}$ dimensions.
We assume in the following that $M \neq 1$.
A straightforward computation then reveals that these spin $s=1$ and $s=3/2$ generators satisfy the (anti-) commutation relations
\begin{align}
[J^{(ij)}_0,  J^{(kl)}_0] &= \delta_{j k} J^{(i l)}_0 + \delta_{i l} J^{(j k)}_0 - \delta_{j l} J^{(i k)}_0 - \delta_{i k} J^{(j l)}_0 \ , \\
[J^{(kl)}_0,  G^p_r] &= \sum_q\rho^{(kl)}_{qp} G^q_{r} \ ,
\end{align}
as well as
\begin{equation}
\{G^p_r, G^q_s\} = 2\, \delta_{pq} L_{r+s}+ \frac{s-r}{2} \left[ \sum_{(kl)} \rho^{(kl)}_{qp}
J^{(kl)}_{r+s} + \nu k \otimes \frac{[\Gamma^p,\Gamma^q]}{2} \right]   \ , \label{eq:hsGG}
\end{equation}
where
\begin{equation}
\rho^{(kl)}_{qp}= \delta_{qk}\delta_{lp}-\delta_{ql}\delta_{kp}\ 
\label{eq:fund_rep}
\end{equation}
are the matrix elements of the fundamental representation of $\so(\mathcal{N})$, while the sum is understood to be over all pairs $(kl)$ (not just the ordered ones).
The Jacobi identities are guaranteed to be satisfied due to  the associativity of the higher spin algebra.

If we now set $\mu = 1/2$, i.\,e. $\nu=0$, we find that the generators defined in \eqref{eq:hsJ}, \eqref{eq:ansatz_G}
indeed satisfy the (anti-)commutation relations of the Lie superalgebra $\mathfrak{osp}( \mathcal{N} |2)$. 
Notice that 
since $k \otimes \frac{[\Gamma^p,\Gamma^q]}{2} $ does not belong to the spin 1 fields of the superalgebra \eqref{eq:hsJ},
it follows from \eqref{eq:hsGG} 
that the $\mathfrak{osp}( \mathcal{N} |2)$ algebra only closes at $\nu=0$ ($\mu=1/2$).  

In other words
we have explicitly constructed a subalgebra $\mathfrak{osp}( \mathcal{N} |2)$ of $\shs_M[\frac{1}{2}]$
with $\mathcal{N}$ as in \eqref{eq:isoodd}.
The relation between the $\mathcal{N}$ and  $M$ is summarized
in the following table
\begin{center}
    \begin{tabular}{| c | c | c |c | c }\hline
		$\shs_2\left[\tfrac{1}{2}\right]$ & $\shs_4\left[\tfrac{1}{2}\right]$& $\shs_8\left[\tfrac{1}{2}\right]$ & $\shs_{16}\left[\tfrac{1}{2}\right]$ &  $\cdots$
    \\[1pt]
	$\mathcal{N}=3$ & $\mathcal{N}=5$ & $\mathcal{N}=7$ & $\mathcal{N}=9$ &  $\cdots$ \\ \hline
    \end{tabular}\ ,
\end{center}

Now let us turn to the case of $\mathcal{N}=2\ell+2$. In that case we set
\begin{equation}
J^{(kl)}_0 = \mathbb{1}\otimes \frac{[\Gamma^k,\Gamma^l]}{4} \ , \qquad J^{(l \mathcal{N})}_0 = \frac{i}{2}\, k \otimes \Gamma^l = - J^{(\mathcal{N} l)}_0 \ ,\\
\label{eq:hsJeven}
\end{equation}
as well as
\begin{equation}\label{eq:hsGeven}
G^{p}_r = \frac{e^{-i\pi/4}}{2}\, \y_\alpha \otimes \Gamma^{p}\ , \qquad G^{\mathcal{N}}_r = \frac{e^{i\pi/4}}{2}\, \y_\alpha k \otimes \mathbb{1} \ ,
\end{equation}
with $1 \leq k,l , p \leq \mathcal{N}-1$. For later use let us also define
\begin{equation}
J^{(kl)}_{\pm} = P_{\pm} \otimes \frac{[\Gamma^k,\Gamma^l]}{4} \ , \qquad J^{(l \mathcal{N})}_{\pm} =  \pm  \frac{i}{2}\, P_{\pm} \otimes \Gamma^l = - J^{(\mathcal{N} l)}_{\pm} \label{jpm} \ .
\end{equation}
Here $\Gamma^p$ are the Gamma matrices from the previous case.
These generators satisfy the same commutation relations as above, together with the following new relation
\begin{equation}
\{G^p_r, G^\mathcal{N}_s\} = \frac{s-r}{2} \bigg[ \sum_{(kl)} \rho^{(kl)}_{\mathcal{N}p} J^{(kl)}_{r+s} + \nu \mathbb{1} \otimes \Gamma^p \bigg]   \ .
\label{eq:GGeven}
\end{equation}
Notice that here $k$ and $l$ take values in the range $1 \leq k,l \leq \mathcal{N}$.
For similar reasons as above we set $\nu=0$, which leads to the following relations between $\mathcal{N}$ and $M$
\begin{center}
    \begin{tabular}{| c| c | c | c |c | c }\hline
		$\shs[\mu]$ &$\shs_2\left[\tfrac{1}{2}\right]$& $\shs_4\left[\tfrac{1}{2}\right]$ & $\shs_{8}\left[\tfrac{1}{2}\right]$ &$\shs_{16}\left[\tfrac{1}{2}\right]$& $\cdots$
    \\[1pt]
	 $\mathcal{N}=2$ &$\mathcal{N}=4$ & $\mathcal{N}=6$ & $\mathcal{N}=8$ &$\mathcal{N}=10$ & $\cdots$ \\ \hline    \end{tabular}\ .
\end{center}


\section{Checking the matrix extended duality}\label{sec:check}

In~\cite{Gaberdiel:2013vva, Creutzig:2013tja} it has been conjectured that the Vasiliev theories based on
the  higher spin algebra $\shs_M[\mu]$ from the previous section are dual to the 't~Hooft limit of the coset CFTs 
\begin{equation}
\frac{\su(N+M)_k \oplus \so(2NM)_1}{\su(N)_{k+M}\oplus \uu(1)_\kappa}
\label{eq:lcoset}
\end{equation}
where $\kappa = MN(M+N)(k+N+M)$, and the 't~Hooft limit is taken by letting
\begin{equation}
N,k\to\infty \qquad \text{with} \qquad \lambda = \frac{N}{k+N} \cong \frac{N}{k+N+M}\qquad \text{held fixed \ .}
\label{eq:tHooft}
\end{equation}
The parameter $\lambda$ in eq.~\eqref{eq:tHooft} is the
 't~Hooft coupling, which has to be identified with the parameter $\mu$ from the higher spin algebra.
In the 't~Hooft limit the central charge of the coset
\begin{equation}
\label{eq:cosetcg}
c = \frac{3 NMk+ (k+N)(M^2-1)}{k+N+M} \cong 3 M \lambda\, k
\end{equation}
diverges proportionally to $N$ which is characteristic of all minimal model dualities.

In the following we first determine the asymptotic symmetry algebra of the Vasiliev higher spin theories, and then compare it to
the chiral symmetry algebra of the coset \eqref{eq:lcoset} in the 't~Hooft limit.

\subsection{Asymptotic symmetry analysis}\label{sec:asym}

Let us begin with a very brief review of the matrix extended Vasiliev higher spin theories. In order to construct these one has to impose a reality condition on $\shs_M[\lambda]$ 
(see \cite{Prokushkin:1998bq, Prokushkin:1998vn}),
which selects the unitary real forms of the subalgebras $\mathfrak{sl}(M)_{\pm}$ at spin $s=1$. Then the full
Vasiliev theory contains the gauge connection 1-forms
$A$, $\bar A$ taking  values in $\shs_M[\mu]$ as well as 0-forms $C$, $\bar C$ taking values in $sB_M[\mu]$.
All fields are defined on a 3-manifold $\mathcal{M}$ with the topology of a solid cylinder.

Since the scalar fields  $C$, $\bar C$ are not relevant for the asymptotic symmetry analysis, we focus in the following on the 
gauge field sector, which can be 
conveniently described in the Chern-Simons formalism that was originally introduced for 3-dimensional (super)gravity \cite{Achucarro:1987vz, Witten:1988hc}
\begin{equation} \label{action1}
I = I_{\rm CS}(A,k_{\rm cs}) - I_{\rm CS}(\bar{A},k_{\rm cs})\ , \qquad
I_{\rm CS} (A,k_{\rm cs})= \frac{k_{\rm cs}}{4 \pi}  \int_{\mathcal{M}} \mbox{Tr}(A \wedge d A + \frac{2}{3} A \wedge A \wedge A)\ .
\end{equation}
Here and in the following we shall employ the same notation as in \cite{Gaberdiel:2013vva, Gaberdiel:2014yla}.
Notice that the components of $A$ (and $\bar A$) contain  $2M^2-1$  spin one gauge fields, and $2M^2$
 gauge fields for each spin $s\geq 3/2$; and one of the spin 2 fields, namely the one corresponding to the conformal
 $\sgl(2)$ subalgebra spanned by the generators defined in \eqref{eq:emt}, corresponds to the graviton.

Next we shall determine the asymptotic symmetry algebra following the known procedure of \cite{Henneaux:2010xg, Campoleoni:2010zq, Gaberdiel:2011wb, Henneaux:1999ib} (see also \cite{Gaberdiel:2012uj} for a review). 
We are interested in gauge connections satisfying the asymptotic AdS$_3$ boundary condition
\begin{equation}
A - A_{\mathrm{AdS}_3} \sim \mathcal{O}(\rho^0)\ ,\quad \bar A - \bar A_{\mathrm{AdS}_3} \sim \mathcal{O}(\rho^0)\ , \qquad \qquad \rho\to \infty\ ,
\label{eq:bc}
\end{equation}
where $\rho$ is the
radial coordinate of the cylinder and  $A_{\mathrm{AdS}_3}$ is the $\mathrm{AdS}_3$  solution given by
\begin{align}\notag
A_{\mathrm{AdS}} &= e^{-\mathbb{L}_0 \rho}\left(\mathbb{L}_1+\frac{\mathbb{L}_{-1}}{4}\right) e^{\mathbb{L}_0 \rho} dx + \mathbb{L}_0 d\rho\ ,\\
  \bar A_{\mathrm{AdS}}&= - e^{\mathbb{L}_0 \rho}\left(\mathbb{L}_{-1}+\frac{\mathbb{L}_{1}}{4}\right) e^{-\mathbb{L}_0 \rho} d\bar x - \mathbb{L}_0 d\rho\ .
\label{eq:absbckr}
\end{align}
Here $x = t/\ell+ \theta$, $\bar x = t/\ell- \theta$, and $t$ is the time coordinate and $\theta$ is the angular coordinate on the cylinder.

The above boundary conditions and further gauge fixing \cite{Campoleoni:2010zq} (see also \cite{Hanaki:2012yf} for the treatment of the supersymmetric case) lead to the following  gauge fixed connection
\be\label{gxcon}
A=b(\rho)^{-1} a(x^+) b(\rho)\ , \qquad b(\rho)=e^{\rho \mathbb{L}_0}\ , 
\ee
with
\be \label{eq:aansatz}
a(x^+)= \mathbb{L}_{1} +\sum_{s,i,j,\varepsilon=\pm} \mathcal{A}^{(s)\,ij\,\varepsilon} (x^+) \, V^{(s)\,ij\,\varepsilon}_{1-s} \ .
\ee
Then the residual gauge symmetry that leaves the form of the gauge fixed connection \eqref{gxcon} unchanged defines the asymptotic symmetry algebra. More concretely, the gauge variation of the connection is described by
\be
\delta_\gamma a=d\gamma+[a,\gamma]=\sum\limits_{s,i,j,m,\epsilon}
c^{(s)\,ij\,\epsilon}_m(a, \gamma) V^{(s) \, ij \, \epsilon}_m\ ,
\ee
where the position dependent (and also field dependent)  gauge transformation parameter $\gamma$ takes values in $\shs_M[\mu]$.
Then imposing the AdS boundary condition and the gauge choice (\ref{eq:aansatz})
imply that
\be\label{cond}
c^{(s)\, ij \, \epsilon}_m=0\ ,\qquad  m\neq 1-s\ ,\quad \forall s,i,j,\epsilon\ .
\ee
Solving the above equations and plugging the results into $\delta _\gamma a=\sum\limits_{s,i,j,\epsilon}c^{(s)\, ij\, \epsilon}_{1-s}V^{(s)\,ij\, \epsilon}_{1-s}$ leads to the variation of the gauge connection under the asymptotic symmetry algebra. We 
can convert the variations into classical operator product expansions (OPEs), which is a convenient form to compare with the CFT computation. We will follow the convention of \cite{Henneaux:1999ib, Gaberdiel:2011wb} 
for this conversion and we shall directly provide the results in the OPE form.

For the purpose of this paper, we are only interested in the OPEs between certain linear combinations of the spin 1 and spin $s=3/2$ generators, namely
those combinations that correspond to the generators of the $\mathfrak{osp}( \mathcal{N} |2)$ subalgebra of 
$\shs_M[\frac{1}{2}]$ from section~\ref{sec:subalg}. In fact, we shall restrict ourselves in the following to the case when $\mathcal{N} =6$ and $M=4$. As far as the comparison with the coset algebra is concerned this choice of generators is of course
as good as any other. It will, however, prove to be a very convenient one in section~\ref{sec:susy} when we study supersymmetry.
Thus, let us observe that the spin 1 generators defined in eq.~\eqref{eq:hsJeven} can be written as
\be
\label{eq:asymJ}
J^{(kl)}_0 = \frac{1}{4} \sum_{i,j} \left( V^{(1)\,ij\,+}_{0} + V^{(1)\,ij\,-}_{0} \right) \bigl( [\Gamma^k,\Gamma^l ] \bigr)_{ij} \ , \quad 
J^{(l\mathcal{N})}_0 = \frac{i}{2} \sum_{i,j} \left( V^{(1)\,ij\,+}_{0} - V^{(1)\,ij\,-}_{0} \right) \bigl( \Gamma^l \bigr)_{ij} \ ,
\ee
where $1\leq k,l \leq \mathcal{N}-1$ and $1 \leq i,j \leq M$, and let us denote by $\mathcal{J}^{(kl)}$ and $\mathcal{J}^{(l\mathcal{N})}$ the corresponding (inverse) linear combinations of the $\mathcal{A}$-fields from eq.~\eqref{eq:aansatz}.
Similarly, we shall denote by $\mathcal{J}_{\pm}^{(kl)}$ and $\mathcal{J}_{\pm}^{(l\mathcal{N})}$ the combinations corresponding to the higher spin fields defined in
\eqref{jpm}. In addition, observe that the spin $s=3/2$ generators defined in eq.~\eqref{eq:hsGeven} can be written as
\begin{align}
\label{eq:asymG}
G^{p}_r &=\frac{e^{-i\pi/4}}{2}\, \sum_{u,v} \left( V^{(3/2)\,uv\,+}_{r} + V^{(3/2)\,uv\,-}_{r} \right) \bigl( \Gamma^p  \bigr)_{uv} \ , \\
G^{\mathcal{N}}_r &= \frac{e^{i\pi/4}}{2}\, \sum_{u} \left( V^{(3/2)\,uu\,+}_{r} - V^{(3/2)\,uu\,-}_{r} \right) \notag  \ ,
\end{align}
where $1\leq p \leq \mathcal{N}-1$ and $1 \leq u,v \leq M$,
and let us denote the corresponding combinations of $\mathcal{A}$-fields by $\mathcal{G}^{p}$ and $\mathcal{G}^{\mathcal{N}}$.
We have computed the classical OPEs between these fields using Mathematica in the case when the matrix factor $M=4$. Our results can be written in the following form:
The classical OPEs of the spin 1  currents $\mathcal{J}$ form an $\mathfrak{so}(\mathcal{N})$ current algebra at  level
\begin{equation}
\frac{2 k_{cs}}{(1+\nu)(1-\nu)} 
\label{eq:lev_hs}
\end{equation}
and their OPE with the $\mathcal{G}$-fields reads
\begin{equation}
\label{eq:ASJG}
\mathcal{J}^{(kl)} (z)  \mathcal{G}^p (w) \sim \frac{\rho^{(kl)}_{qp} \mathcal{G}^q (w)}{z-w} \ ,
\end{equation}
while the OPEs between spin-3/2 operators are given by
\begin{multline}
\mathcal{G}^p(z)\mathcal{G}^q(w)\sim
\frac{\delta_{p q} 4 k_{\mathrm{cs}}}{(z-w)^3} + \frac{\delta_{p q}2\,\mathcal{T}(w)}{z-w} +\frac{\mathcal{V}^{pq}(w)}{z-w} \\
{}+ \frac{ 2 \left( (1 + \nu) \mathcal{J}_+^{(p q)} + (1 - \nu) \mathcal{J}_-^{(p q)} \right) (w)  }{(z-w)^2}
+ \frac{  \partial \left( (1 + \nu) \mathcal{J}_+^{(p q)} + (1 - \nu) \mathcal{J}_-^{(p q)} \right) (w)  }{z-w}  \ .
\label{eq:ASGG}
\end{multline}
Here the indices $k,l,p,q$ take values in the range $1 \leq k,l,p,q\leq \mathcal{N}$, and $\mathcal{T}$ is the energy momentum tensor given by the expression
\begin{equation}
\mathcal{T} = \frac{1}{2 M} \sum_{i,\varepsilon} \mathcal{A}^{(2)\,ii\,\varepsilon} + \frac{1}{2 k_{\mathrm{cs}}} \sum_{i,j} \left( (1 + \nu) \mathcal{A}^{(1)\,ij\,+} \mathcal{A}^{(1)\,ij\,+} + 
(1 - \nu) \mathcal{A}^{(1)\,ij\,-} \mathcal{A}^{(1)\,ij\,-} \right)
\end{equation}
with central charge
\begin{equation}
c = 6 k_{\mathrm{cs}}\ .
\label{eq:cc_hs}
\end{equation}
The operator $\mathcal{V}^{pq}$ contains all the terms of the first order pole in the 
 OPE between $\mathcal{G}^p(z) \mathcal{G}^q(w)$, which are not part of the energy momentum tensor and are not
derivatives of the currents. It is given as
\begin{multline}
\label{eq:vas}
\mathcal{V}^{p q} :=  - \frac{(1+ \nu)(1- \nu)}{2 k_{\mathrm{cs}}} \Biggl[\delta_{p q} \sum_{i,j} \left( \mathcal{A}^{(1)\,ij\,+} \mathcal{A}^{(1)\,ij\,+} + \mathcal{A}^{(1)\,ij\,-} \mathcal{A}^{(1)\,ij\,-} \right) \\ {}+ 
\sum_{i,j,k,l} C^{(pq)}_{i j k l} \mathcal{A}^{(1) \,  i j \, -} \mathcal{A}^{(1) \, kl \, +} \Biggr]
\end{multline}
with
\begin{equation*}
C^{(pq)}_{i j k l} = \left( \Gamma^p_{i l } \Gamma^q_{k j} + \Gamma^q_{i l } \Gamma^p_{k j} \right)\,,\qquad C^{(p \mathcal{N})}_{i j k l} = - i \left( \Gamma^p_{i l } \delta_{k j} - \delta_{i l } \Gamma^p_{k j} \right)
\,,\qquad C^{(\mathcal{N} \mathcal{N})}_{i j k l} = 2 \delta_{i l } \delta_{k j}\,.
\end{equation*}
Let us remark that if we set $\nu=0$ and ignore the $\mathcal{N}^{\mathrm{th}}$ supercharge we recover (the structure of) the expression in eq.~(6.20) of \cite{Creutzig:2014ula}.
Now from the above results we can conclude that at $\nu = 0$ the wedge modes of the $\mathcal{J}$ and $\mathcal{G}$ generators 
form an $\osp(\mathcal{N}|2)$ subalgebra. However, as we shall explain in section~\ref{sec:susy} it also follows from these results that
the asymptotic symmetry algebra does not possess extended superconformal supersymmetry.
But before we address this point, we will compare these results with the OPEs computed from the coset CFT.

\subsection{The dual coset theories} \label{sec:coset}

In this section we shall define the coset currents corresponding to the asymptotic symmetry generators defined in eqs.~\eqref{eq:asymJ}, \eqref{eq:asymG} and \eqref{eq:aansatz}, and compute the (quantum) OPEs between
them. In contrast to section~\ref{sec:asym}, we will obtain the results for all values of the parameter $M$ with $M=2^{\mathcal{N}/2-1}$ for $\mathcal{N}$ even.
In the next section we will then show that these results (for $M=4$) agree in the 't~Hooft limit with the classical OPEs that we obtained in eqs.~\eqref{eq:ASJG} and \eqref{eq:ASGG}.

In the following we shall employ the same notation as in \cite{Candu:2013fta}.
First, let us introduce a basis for the currents corresponding to the factor
$\mathfrak{su}(N+M)_k$ in the numerator of the coset \eqref{eq:lcoset}. It is convenient to choose a basis that splits according to the decomposition
\begin{equation}
\su(N+M)_k\simeq \underbrace{\su(N)_k}_{J^A}\oplus\underbrace{\su(M)_k}_{J^I}\oplus \underbrace{\uu(1)}_{J}\oplus \underbrace{(N,\bar M)_{N+M}}_{J^{ai}}\oplus \underbrace{(\bar N ,M)_{-N-M}}_{\bar{J}^{ai}}
\label{eq:dec_sunm}
\end{equation}
where the lower index denotes the $J$-charge. Thus, $J^A$, $J^I$ and $J$ denote the basis elements generating the affine superalgebras $\su(N)_k$, $\su(M)_k$ and $\uu(1)$,
while $J^{ai}$ and $\bar{J}^{ai}$ are the elements transforming in the representation $(N,\bar M)$ and $(\bar N ,M)$, respectively.

Then there are the currents that can be constructed from the $NM$ Dirac fermions $\psi^{ai}$ and their conjugates $\bar{\psi}^{ai}$
corresponding to the $\so(2NM)_1$ factor of the coset. A basis for these are given by
\begin{equation}
K^{A}= t^A_{ab}:\psi^{ai}\bar\psi^{bi}:\ ,\qquad K^{I}= t^{I}_{ij}: \bar \psi^{a i}\psi^{aj}:\ ,\qquad  K =\psi^{ai}\bar\psi^{ai}\ ,
\label{eq:twosuM}
\end{equation}
where $t^A_{ab}$, $t^I_{ij}$ are the matrix elements of the fundamental representation of the algebras $\mathfrak{su}(N)$ and $\mathfrak{su}(M)$ generated by
the currents $J^A$ and $J^I$, see Appendix A.
These currents generate the algebra
\begin{equation}
\underbrace{su(N)_M}_{K^{A}} \oplus \underbrace{\su(M)_N}_{K^{I}} \oplus \underbrace{\uu(1)_{NM}}_{K} \ ,
\end{equation}
and with respect to them the Dirac fermions $\psi^{ai}$ transform in the representation $(N,\bar M)_{1}$, and their
conjugates $\bar \psi^{ai}$ in the representation $(\bar N,M)_{-1}$. All the OPEs between the currents defined in \eqref{eq:dec_sunm} and \eqref{eq:twosuM} can be found in appendix A.

Finally, we have to specify how the denominator  of the coset $\su(N)_{k+M}\oplus \uu(1)_\kappa$ is embedded into the numerator.
The embedding is given by
\begin{equation}
\tilde J^{A}:= J^{A} + K^{A}\ ,\qquad \tilde J:= J + (N+M)K\ ,
\label{eq:embsuN}
\end{equation}
where the relative coefficient between $J$ and $K$ in the definition of $\tilde J$ is determined in such a way that $J^{ai}$ and $\psi^{ai}$ have the same $\tilde J$-charge,
which is required by the $\mathcal{N}=1$ supersymmetry of the parent coset theory of the coset~\eqref{eq:lcoset} (see e.\,g.\ \cite{Candu:2013fta} for more information).
The overall factor in $\tilde J$ is chosen such that all representations have integer $U(1)$ charges. The corresponding
 OPE is given by
\begin{equation}
\tilde J(z)\tilde J(w)\sim  \frac{\kappa}{(z-w)^2}\ ,\qquad \kappa:= NM(N+M)(k+N+M)\ .
\label{eq:levu1}
\end{equation}

After this preliminary work, we can now define the coset~\eqref{eq:lcoset} as the algebra of normal ordered differential polynomials in the numerator currents
that are regular with respect to the denominator currents of eq.~\eqref{eq:embsuN}.
A particular example of a coset current is, of course, the energy momentum tensor. It can be computed according to the GKO construction \cite{Goddard:1986ee, god}
as the difference between the energy momentum tensor of the numerator and the denominator of the coset, the result being
\begin{multline}
T = \frac{1}{2(k+N+M)} \left[(J^IJ^I) + (K^I K^I) +(J^{ai}\bar J^{ai})+(\bar J^{ai}J^{ai})- 2JK/NM \right.\\
\left.{}-2K^A J^A-k :(\psi^{ai}\partial\bar \psi^{ai}+\bar \psi^{ai}\partial \psi^{ai}):\right] \ .
\label{eq:cosetT_expl}
\end{multline}

Other currents of low spin can be written down explicitly as well. In particular, it is easy to convince ourselves that all the $2M^2-1$  spin 1 fields of the coset~\eqref{eq:lcoset}
are given by
\begin{equation}
J^I\ ,\qquad  K^I\ ,\qquad  U= \frac{J - k K}{k+N+M}\ ,
\label{eq:coset_s1c}
\end{equation}
and the $2M^2$ spin $s=3/2$ currents by
\begin{equation}
\psi^{ai} \bar{J}^{aj}\ ,\qquad  J^{a i} \bar{\psi}^{a j}\ ,
\end{equation}
where $i,j$ are free indices.
In \cite{Candu:2013fta} it is explained in some detail how currents of spin $s\geq 2$ can be constructed.
However, for our purposes it suffices to consider particular combinations of spin 1 and spin $s=3/2$ currents --- namely the ones corresponding to the generators 
defined in eqs.~\eqref{eq:asymJ}, \eqref{eq:asymG} and \eqref{eq:aansatz} from the previous section.
These are given by
\begin{equation}
J^{(kl)} =  \frac{\left( J^{I} + K^{I} \right) \tr  \left( t^I [\Gamma^k,\Gamma^l] \right)}{4} \ , \qquad
J^{(l \mathcal{N})} = \frac{i \left( J^{I} - K^{I} \right) \tr \left( t^I \Gamma^l \right)}{2} \ , \\
\label{eq:coset_js}
\end{equation}
and $J^{(\mathcal{N} l)} := - J^{(l \mathcal{N})}$, as well as
\begin{equation}
G^p = \frac{\left( J^{a i} \bar{\psi}^{a j} + \psi^{ai} \bar{J}^{aj}\right) \Gamma^p_{ji}}{\sqrt{k+N+M}} \ , \qquad
G^\mathcal{N} = \frac{i \left( J^{a i} \bar{\psi}^{a i} - \psi^{ai} \bar{J}^{ai} \right)}{\sqrt{k+N+M}} \ ,
\label{eq:coset_gs}
\end{equation}
where the free indices $k,l,p$ take values in the range $1 ,2,3, \dots , \mathcal{N}-1$ and $\mathcal{N}$ is even. The Gamma matrices 
$\Gamma^p$ (with $p = 1, \dots, \mathcal{N}-1$) are realized in $M=2^{\mathcal{N}/2-1}$ matrix dimensions, 
and satisfy the Clifford algebra
 relation~\eqref{eq:cliff}.

Then a straightforward computation reveals that the OPE between the spin 1 currents~\eqref{eq:coset_js} form an $\mathfrak{so}(\mathcal{N})$ current algebra at level
\begin{equation}
\frac{(k+N)M}{4}
\label{eq:lev_cos}
\end{equation}
and their OPE with the spin $s=3/2$ currents \eqref{eq:coset_gs} are given by
\begin{equation}
\label{eq:cosetJG}
J^{(kl)} (z)  G^p (w) \sim \frac{\rho^{(kl)}_{qp} G^q (w)}{z-w} \ .
\end{equation}
Here the indices $k,l,p$ are allowed to take values in the range $1 ,2,3, \dots , \mathcal{N}$ and $\rho^{(kl)}_{qp}$ are the matrix elements of the fundamental representation of $\so(\mathcal{N})$.

Finally, we can also compute the OPEs between the spin $s=3/2$ generators and themselves. The results of this computation read
\begin{align}
G^p(z)G^q(w) &\sim
\frac{\delta_{p q}b}{(z-w)^3} + \frac{\delta_{p q}2\,T(w)}{z-w} +\frac{V^{pq}(w)}{z-w} \label{eq:genOpe_GGpq} \\
 &\quad {}+ \gamma \left[\frac{ (N J^{I} + k K^{I}) (w)  }{(z-w)^2} + \frac{  \partial ( N J^{I} + k K^{I})(w)  }{2 (z-w)}\right]  \tr  \left( t^I [\Gamma^p,\Gamma^q] \right) \ ,
 \notag \\ \notag
 G^p(z)G^\mathcal{N}(w) &\sim
 \frac{V^{p \mathcal{N}}(w)}{z-w}+2 i \gamma \Biggl[\frac{(N J^{I} - k K^{I})(w) }{(z-w)^2}+ \frac{ \partial ( N  J^{I} - k K^{I})(w)}{2 (z-w)}\Biggr]   \tr  \left( t^I \Gamma^p \right) \ , \\
G^\mathcal{N}(z)G^\mathcal{N}(w) &\sim \frac{b}{(z-w)^3} + \frac{2\,T(w)}{z-w} +\frac{V^{\mathcal{N} \mathcal{N}}(w)}{z-w} \ , \notag
\end{align}
where $p,q = 1 ,2,3, \dots , \mathcal{N}-1$ and
\begin{equation}
b=\frac{2kN M}{k+N+M}
 \ , \qquad
\gamma= \frac{1}{k+N+M} \label{eq:sol_b} \ ,
\end{equation}
while the components of the operator $V$ are given by
\begin{equation}
V^{pq} =  - \gamma  \left[ \delta^{pq} (J^{I}J^{I}) + \delta^{pq} ( K^{I}K^{I} )
+J^I K^J \tr\left( \Gamma^p t^I \Gamma^q t^J + \Gamma^q t^I \Gamma^p t^J \right) \right]\ ,
\label{eq:vpq}
\end{equation}
as well as
\begin{equation}
V^{p\mathcal{N}} = i\,\gamma J^I K^J \tr\left( \Gamma^p t^I t^J - t^I \Gamma^p t^J \right) \ , \qquad
V^{\mathcal{N}\mathcal{N}} =  - \gamma \left( J^{I} + K^I \right)^{2}  \ .
\label{eq:vpmvmm}
\end{equation}

\subsection{Matching the symmetry algebras}

In the previous two sections we have computed classical OPEs between operators of low spins in the asymptotic symmetry algebra 
based on $\shs_M[\mu]$ as well as
`quantum' OPEs in the dual coset CFT. Now we shall check whether these results agree in the 't Hooft limit, cf.\ \eqref{eq:tHooft}.

The OPEs between the $\mathfrak{so}(\mathcal{N})$ currents of spin 1 agree if the central terms~(\ref{eq:lev_hs}, \ref{eq:lev_cos}) match.
This fixes the relation between $k_{cs}$ and $k$ in the 't Hooft limit as follows
\begin{equation}
k_{cs} = \frac{k M \lambda }{2}\ . 
\label{eq:lev_rel}
\end{equation}
Next, the OPEs~(\ref{eq:ASJG}, \ref{eq:cosetJG}) between the currents of spin 1 and 3/2 are manifestly the same, while
the central charges~(\ref{eq:cosetcg}, \ref{eq:cc_hs}) of the Virasoro algebras also agree using \eqref{eq:lev_rel}.
It only remains to consider the OPE between the spin $s=3/2$ operators, 
see eqs.~\eqref{eq:ASGG} and \eqref{eq:genOpe_GGpq}, which we do next.

The higher spin counterpart of the central term $b$  appearing in the coset OPE \eqref{eq:genOpe_GGpq}  is  $4 k_{cs}$, see eq.~\eqref{eq:ASGG},
and this agrees (in the 't Hooft limit) with the definition~\eqref{eq:sol_b} if one uses \eqref{eq:lev_rel}.

Next, let us consider the second order pole. E.\,g.\ from eqs.~\eqref{eq:ASJG} and \eqref{eq:cosetJG} it follows that we have to identify the spin 1 currents
$\mathcal{J}^{(kl)}_{+}$ and $\mathcal{J}^{(kl)}_{-}$ of the asymptotic symmetry algebra with the coset currents 
$J^I \tr  \left( t^I [\Gamma^k,\Gamma^l] \right)/4$ and $K^I \tr  \left( t^I [\Gamma^k,\Gamma^l] \right)/4$ for $k,l \leq \mathcal{N}-1$, and $i J^I \tr \left( t^I \Gamma^k \right)/2$ 
and $-i K^I \tr  \left( t^I \Gamma^k \right)/2$ for  $l=\mathcal{N}$, respectively. 
Their coefficients agree
\begin{equation} \label{eq:nucomp}
1 + \nu \cong  \frac{2 N }{k+N+M} \quad \mathrm{and} \quad 1 - \nu \cong  \frac{2 k }{k+N+M}
\end{equation}
due to eq.~\eqref{eq:tHooft} and the relation $\nu = 2 \lambda -1$.

Finally, consider the terms in the first order pole. Because
the coefficients of the derivatives of the spin 1 currents and the energy momentum tensor agree due to eq.~\eqref{eq:nucomp},
it remains to consider only the operators $\mathcal{V}^{pq}$ and $V^{pq}$, respectively. 
Naively, 
one would expect that their contribution to the OPE vanishes in the 't~Hooft limit due to the overall 
factor $1/(k+N+M)$. However, as explained e.\,g.\ in \cite{Candu:2013uya} taking the 't~Hooft limit of $\mathcal{W}$-algebras
involves a rescaling of the (simple) operators and if one carries out this analysis correctly, 
one finds that the operator does contribute in the limit. 
The comparison of these bilinear terms provides a non-trivial consistency check for the
holographic duality proposed in \cite{Gaberdiel:2013vva, Creutzig:2013tja}.

Now in order to see whether the operators $\mathcal{V}^{pq}$ and $V^{pq}$ do indeed agree, observe that we have to identify the (traceless part of
the) operators 
$\mathcal{A}^{(1) \, kl \, +}$ and $\mathcal{A}^{(1) \, kl \, -}$ with 
$J^I t^I_{kl}$ and $K^I  t^I_{kl}$. In addition, we need that
\begin{equation}
\frac{(1 + \nu) (1 - \nu)}{2 k_{cs}} = \frac{(1 + \nu) (1 - \nu)M}{8 k_{cs}} \cong \frac{1}{k+N+M} \ ,
\end{equation}
which works due eq.~\eqref{eq:lev_rel}, where we have used the fact that we have set $M=4$ in section~\ref{sec:asym}.

In conclusion, we have compared in this section non-trivial OPEs between operators
of spin $s=3/2$ in the asymptotic symmetry algebra of $\shs_M[\mu]$  and the dual coset chiral algebra
and found agreement. In particular, we have shown that the bilinear terms appearing in the first order pole
of these OPEs precisely agree in the 't~Hooft limit. 


\section{Existence of extended supersymmetries}\label{sec:susy}

In this section we shall study the structure of the asymptotic symmetry algebra and chiral algebra of the cosets~\eqref{eq:lcoset} more systematically. 
Our objective is to determine 
whether the asymptotic symmetry algebra of the matrix extended higher spin theory, or equivalently the coset chiral algebra, is supersymmetric.

Let us try to get some hint on how to proceed from the discussion in section~\ref{sec:subalg}, where we have shown that the higher spin 
algebra $\shs_M[\mu]$ contains for $\mu = 1/2$ and
\begin{equation}
M=2^{\mathcal{N}/2-1} \quad \mathrm{with} \ \mathcal{N} \ \mathrm{even} \ , 
\end{equation}
the Lie superalgebra $\osp(\mathcal{N}|2)$ as a subalgebra. Based on this observation, 
we might expect the asymptotic symmetry algebra of $\shs_M[\mu]$
to exhibit $\mathcal{N}$-extended superconformal symmetry at $\mu = 1/2$. However, as we have already noted at the end of section~\ref{sec:asym},
it follows from our analysis of low lying OPEs in sections\ \ref{sec:asym} and \ref{sec:coset} that this is not the case. The $\mathcal{N}$-extended superconformal symmetry is broken by bilinear terms.
In the following let us explain this point in some more detail  from the coset point of view, where the condition that $\mu = 1/2$ corresponds to setting
\begin{equation}
k=N \ .
\label{eq:restriction}
\end{equation}
First, observe that for $k=N$ the OPEs between the 
spin  $s=3/2$  operators computed in eq.~\eqref{eq:genOpe_GGpq} simplify considerably and take the form
\begin{equation}
G^p(z)G^q(w)\sim 
\frac{\delta_{p q}b}{(z-w)^3} + \frac{\delta_{p q}2\,T(w)}{z-w} +\frac{V^{pq}(w)}{z-w} +  
\sigma \rho^{(kl)}_{qp}\left[\frac{J^{(kl)}(w)}{(z-w)^2} + \frac{\partial J^{(kl)}(w)}{2(z-w)}\right]  \ ,
\label{eq:ope_GGkN}
\end{equation}
where $\sigma = -2 N/(2 N+M)$. Recall that the operator $V^{pq}$ contains all the terms of the first order pole in the 
$G^p(z)G^q(w)$ OPE which are not part of the energy momentum tensor and are not
derivatives of the currents. The explicit expression can be found in eqs.~\eqref{eq:vpq} and \eqref{eq:vpmvmm}.
Then the existence of supersymmetry is paramount to testing whether the operator $V^{pq}$ can be written in terms of bilinears of the spin 1 currents 
from eq.~\eqref{eq:coset_js}. Note that it is plausible that this can be done since $V^{pq}$ is bilinear in the coset currents $J^I$ and $K^I$ and contains none of the $2M^2$ spin 2 coset currents.
However, we have checked that $V^{pq}$ can be rewritten
solely in terms of the $\so(\mathcal{N})$ currents~\eqref{eq:coset_js} only if $M=1$ or $M=2$.

The fact that the coset algebra is supersymmetric for $M=1,2$ is of course well-known. The case $M=1$ corresponds to the 
$\mathbb{C}P^{N}$ Kazama-Suzuki cosets,  which are $\mathcal{N}=2$ supersymmetric, see \cite{Kazama:1988qp}.
The reason why the operator $V^{pq}$ vanishes identically in this case is actually that there are no coset currents $J^I$ and $K^I$. The case $M=2$, on the other hand, corresponds to the Wolf space cosets, 
which are known to have non-linear large $\mathcal{N}=4$ supersymmetry for any $N$ and $k$ \cite{Spindel:1988sr, VanProeyen:1989me, sevrin}.
In this case the operator $V_{pq}$ is a bilinear expression in the $\mathfrak{g}=\so(4)$ currents  of the large (non-)linear $\mathcal{N}=4$ superconformal algebra simply because this current algebra
contains all the coset currents $J^I$, $K^I$, see also \cite{Gaberdiel:2013vva}. Note that nothing is special at $\mu = 1/2$ (i.\,e. $k=N$) because
$V_{pq}$ has the same form everywhere.

Let us emphasize that from the above analysis we cannot draw the conclusion that the asymptotic symmetry algebra and the coset cannot possess any supersymmetry for $M\geq2$. We have merely shown that the most naive ansatz 
for the spin 1 currents and the `supercharges' does not yield an $\mathcal{N}$-extended superconformal subalgebra of the asymptotic symmetry and coset algebras except for $M=1$ and $M=2$. However, 
we have checked that even a much more general ansatz does not provide  $\osp(\mathcal{N}|2)$ supersymmetry.
Again let us present our results in the coset picture. For the spin 1 currents we make the ansatz
\begin{equation}
L^{P} = C^{PI}_+J^{I}+C^{PI}_{-} K^{I}\ ,
\label{eq:Lcurrents}
\end{equation}
and for the `supercharges' we set
\begin{equation}
G^p = \frac{1}{\sqrt{k+N+M}} (x_+^p J^{a i} \bar{\psi}^{a j} + x_-^p \psi^{ai} \bar{J}^{aj} )\Gamma^p_{ji}\ .
\label{eq:coset_ascgeneral}
\end{equation}
Notice that trilinears in the fermions cannot appear due to e.\,g.\ the neutrality conditions with respect to the $\uu(1)$ factor in the denominator of the coset.
Here $C^{PI}_{\pm}$, $x_\pm^p$ and $\Gamma^p$ are unknown constants and matrices, respectively. The coset is supersymmetric if and only if these unknowns can be chosen in such a way that the currents $L^P$
generate  a subalgebra $\hat{\mathfrak{g}}\subset \su(M)_k\oplus\su(M)_N$ and the
OPEs between the supercharges take the form
\begin{multline}
G^p(z)G^q(w)\sim \frac{\delta_{p q} \tilde{b}}{(z-w)^3} + \frac{\delta_{p q}2T(w)}{z-w} +\frac{\tilde{V}_{pq}(w)}{z-w}+\\
{}+  \tilde{\sigma} \rho^P_{qp}\left[\frac{L^P(w)}{(z-w)^2} + \frac{\partial L^P(w)}{2(z-w)}\right]  + \xi\, \rho^U_{qp}\left[\frac{U(w)}{(z-w)^2} + \frac{\partial U(w)}{2(z-w)}\right]\ ,
\label{ope_GG}
\end{multline}
\emph{with $\tilde{V}_{pq}$ being a quasiprimary operator bilinear in the currents $L^P$}. Notice that
the  Jacobi identities $LGG$ and $UGG$
imply that the (a priori unknown) coefficients $\rho^P_{qp}$ and $\rho^U_{qp}$ in eq.~\eqref{ope_GG} are precisely the matrix elements of 
the orthogonal representations of $\hat{\mathfrak{g}}$ and $\uu(1)$, respectively, 
in which the supercharges transform under the action of the currents $L^P$ and $U$, i.e.\
\begin{equation}
L^{P} (z) G^p (w) \sim \frac{\rho^{P}_{qp} G^q (w)}{z-w} \ ,\quad U (z) G^p (w) \sim \frac{\rho^{U}_{qp} G^q (w)}{z-w}\ .
\label{eq:opeLG}
\end{equation}
After some algebra, we find the following solution: The supercharges split into two sets indexed by
\begin{equation}
\Pi'= \{p'\mid x_+^{p'} =x_-^{p'}=  1\}\ ,\quad \Pi''= \{p''\mid x^{p''}_+= - x_-^{p''} = 1\}\ .
\label{eq:2sets}
\end{equation}
where the corresponding matrices $\{\Gamma^{p'}\}$
and $\{\Gamma^{p''}\}$ generate two mutually commuting Clifford algebras
\begin{equation}
\{\Gamma^{p'},\Gamma^{q'}\} = 2\delta^{p'q'}\mathbb{1}_M\ ,\quad \{\Gamma^{p''},\Gamma^{q''}\} = -2\delta_{p''q''}\mathbb{1}_M\ ,\quad [\Gamma^{p'},\Gamma^{q''}] = 0\ .
\label{eq:2cliffords}
\end{equation}
The constants $C^{PI}_{\pm}$ appearing in the spin 1 currents are given by the solutions of the relation
$t^P_\pm = C^{PI}_\pm t^I$, where
\begin{equation}
t^{ (p'q')}_\pm = \frac{u_\pm}{4} [\Gamma^{p'},\Gamma^{q'}]\ ,\quad t^{(p''q'')}_\pm = \frac{v_\pm}{4} [\Gamma^{p''},\Gamma^{q''}]\ ,\quad t^{(p'q'')}_{\pm} =-t^{(q''p')}_{\pm} = \frac{w_\pm}{4} \{\Gamma^{p'},\Gamma^{q''}\}\ .
\label{eq:basistw}
\end{equation}
Clearly, $t^{(p'q')}_\pm$ generate $\so(\mathcal{N}')$, where $\mathcal{N}'$ is the cardinality of $\Pi'$,  $t^{(p''q'')}_\pm$ generate $\so(\mathcal{N}'')$, where $\mathcal{N}''$ is the cardinality of $\Pi''$,
while $t^{(p'q'')}_\pm$ transform in the representation $(\mathcal{N}',\mathcal{N}'')$ of $\so(\mathcal{N}')\oplus\so(\mathcal{N}'')$.
A direct computation shows that altogether the generators~\eqref{eq:basistw} satisfy the commutation relations of $\so(\mathcal{N})$, 
where $\mathcal{N}=\mathcal{N}'+\mathcal{N}''$.

Now there are two cases to consider: 1) the matrices~\eqref{eq:basistw} are traceless or 2) they are non-traceless.
In fact, it is easy to conclude from the commutation  relations of the generators~\eqref{eq:basistw} that these cases correspond to 1) $(\mathcal{N}',\mathcal{N}'')\neq (1,1)$
and 2) $\mathcal{N}'=\mathcal{N}''= 1$. For $\tilde{b}$ and $\tilde{\sigma}$ we get the same values as above, see eq.~\eqref{eq:sol_b} and 
the expression below \eqref{eq:ope_GGkN}.
  
Let us first consider case 2). This case corresponds to the $\mathcal{N}=2$ Kazama-Suzuki cosets introduced before.
Concretely, since there are no currents $L^P$, only the $U$ current will contribute to the second pole of the OPE~\eqref{ope_GG}.
The most general solution is given by 
\begin{equation}
 x^1_\pm=1\ ,\quad x^2_\pm = \pm 1 \ ,\quad \Gamma^1 = i \Gamma^2\ ,\quad \gamma=2\ ,\quad \rho^U=\begin{pmatrix} 0 & -i\\ i & 0\end{pmatrix}\ ,
\label{eq:case2}
\end{equation}
where $\Gamma^1$ is any matrix that squares to identity.

In case 1), corresponding to $(\mathcal{N}',\mathcal{N}'')\neq (1,1)$ we have that $\xi=0$, and,
consequently,  the current $U$ will not appear in the second pole of the OPE~\eqref{ope_GG}. In addition, it can be shown that the solution exists if and only if $$k=N.$$
This solutions can be thought of as a generalization of the previous ansatz (with $M > 1$). Indeed, if choose $\mathcal{N}'' =1$, $\Gamma^{p''} = i \mathbb{1}$ and an irreducible representation for
$\{\Gamma'\}$,  we obtain the old ansatz.
Of course, we can choose reducible representations of the two Clifford algebras to get smaller values of $\mathcal{N}\leq 2\ell+2$ for fixed $M=2^\ell$.

We have checked that also the more general solutions do not yield supersymmetry either except for $M=1,2$. In fact, the operator $\tilde{V}_{pq}$ is given by
\begin{equation}
\tilde{V}_{pq} =  - \frac{1}{k+N+M}  \left[ \delta^{pq} (J^{I}J^{I}) + \delta^{pq} ( K^{I}K^{I} )+J^I K^J\tr(x^p_+x^q_- \Gamma^pt^I\Gamma^qt^J + x^p_-x^q_+ \Gamma^q t^I\Gamma^p t^J) \right]\ ,
\label{eq:vpqgeneral}
\end{equation}
and we have checked that it cannot be rewritten solely in terms of the currents $L^p$ appearing in the second order pole of the OPE between the spin $s=3/2$ operators.


\section{Conclusions} \label{sec:concl}


In this paper we have performed the asymptotic symmetry analysis for the $\shs_M[\mu]$ Vasiliev theories for $M > 2$, and used the 
results to test the holographic duality \cite{Creutzig:2013tja} between these theories and the 't~Hooft limit of the cosets \eqref{eq:lcoset}.
In particular, we have computed OPEs between operators of low spin ($s \leqslant 2$)   
in the asymptotic symmetry algebra of $\shs_4[\mu]$, and compared these to the OPEs 
between the corresponding operators of the dual coset theory and found agreement. 
Our computations reproduce similar results  obtained by different methods in \cite{Creutzig:2013tja, Creutzig:2014ula}.

We have also studied whether the asymptotic symmetry algebra and the coset chiral algebra exhibit (extended) supersymmetries
for $M > 2$ .
Based on the observation that the higher spin algebra $\shs_M[\mu]$ at $\mu = 1/2$ and $M=2^{\mathcal{N}/2-1}$ with $\mathcal{N}$ even
contains the Lie superalgebra $\osp(\mathcal{N} \vert 2)$ as a subalgebra, we have
made an ansatz for the supercharges in the asymptotic symmetry algebra of $\shs_M[\frac{1}{2}]$
as well as in the dual cosets, see eqs.~\eqref{eq:hsGeven}, \eqref{eq:aansatz} and \eqref{eq:coset_gs}. Then we have computed the OPEs between these operators and found
that they contain bilinear terms built out of spin 1 currents that are not part of the superconformal algebra, cf.\ \eqref{eq:vas}, \eqref{eq:vpq}, \eqref{eq:vpmvmm}, and hence break superconformal supersymmetry. 
We have checked that
these terms are also present for more general ans\"atze for the supercharges. 

In addition, we have studied the relation between the matrix extended higher spin theories and 
the Vasiliev theories based on the higher spin algebra $\shs^E\! \left( \mathcal{N} \vert 2, \mathbb{R} \right)$,
which have played a role in the recent works of \cite{Henneaux:2012ny} and \cite{Creutzig:2014ula}.
We have shown that the higher spin algebra $\shs^E\! \left( \mathcal{N} \vert 2, \mathbb{R} \right)$ 
is isomorphic to $\shs_M[\frac{1}{2}]$ with $M=2^\ell$ when $\mathcal{N}=2\ell+2$, and
$\left( \shs_M[\frac{1}{2}] \right)^{\mathbb{Z}_2}$ with $M=2^l$ when $\mathcal{N}=2\ell+1$.
This implies that we have found the CFT duals of the $\shs^E\! \left( \mathcal{N} \vert 2, \mathbb{R} \right)$
Vasiliev theories in the case when $\mathcal{N}$ is even, thereby complementing the dualities conjectured
in \cite{Creutzig:2014ula}. For even $\mathcal{N}$ the dual theories are given simply as the 't~Hooft limit of the
cosets in \eqref{eq:lcoset}.
In addition, it follows immediately from the above discussion that the asymptotic symmetry algebra of $\shs^E\! \left( \mathcal{N} \vert 2, \mathbb{R} \right)$
does \emph{not} contain the $\mathcal{N}$-extended superconformal algebra as a subalgebra, for both the case of $\lambda$-deformed and undeformed algebras.\footnote{We have confirmed this also by a direct computation in the asymptotic symmetry algebra of $\shs^E\! \left( \mathcal{N} \vert 2, \mathbb{R} \right)$.}


\section*{Acknowledgments}
We thank Matthias Gaberdiel for useful discussions and guidance and Yasuaki Hikida and Peter R\o{}nne for comments. This work is partially supported by the Swiss National
Science Foundation and the NCCR SwissMAP.


\appendix

\section{OPEs between coset currents} \label{sec:cosetopes}

In this appendix we provide explicit expressions for the operator product expansions (OPEs) between currents of the coset~\eqref{eq:lcoset}.
As in section~\ref{sec:coset}, we shall employ the same notation as in \cite{Candu:2013fta}.

Let us begin with the OPEs between the currents that correspond to the $\su(N+M)_{k}$ coset factor. 
In the basis  given in \eqref{eq:dec_sunm} these take the form 
\begin{align}\label{eq:sumn_OPE}
J^A(z)J^B(w)&\sim \frac{k \delta_{AB} }{(z-w)^2}+\frac{f_{ABC}J^C(w)}{z-w}\ ,\\ \notag
J^I(z)J^J(w)&\sim \frac{k \delta_{IJ} }{(z-w)^2}+\frac{{f_{IJK}} J^K(w)}{z-w}\ ,\;
 J(z)J(w)\sim \frac{kNM(N+M)}{(z-w)^2}\ ,\\ \notag
J^A(z)J^{ai}(w)&\sim \frac{t^A_{ba}J^{bi}(w)}{z-w}\ ,\; J^I(z)J^{ai}(w)\sim \frac{-t^I_{ij}J^{aj}(w)}{z-w}\ ,\; J(z)J^{ai}(w)\sim \frac{(N+M)J^{ai}(w)}{z-w}\ ,\\ \notag
J^A(z)\bar J^{ai}(w)&\sim \frac{-t^A_{ab}\bar J^{bi}(w)}{z-w}\ ,\;
J^I(z)\bar J^{ai}(w)\sim \frac{t^I_{ji}\bar J^{aj}(w)}{z-w}\ ,\;  J(z)\bar J^{ai}(w)\sim \frac{-(N+M)\bar J^{ai}(w)}{z-w}\ ,\\ \notag
 J^{ai}(z)\bar{J}^{bj}(w)&\sim \frac{k \delta_{ij}\delta_{ab}}{(z-w)^2} + \frac{\delta_{ij}t^A_{ba}J^A(w)-\delta_{ab}t^I_{ij}J^I(w)+\frac{1}{NM}\delta_{ij}\delta_{ab}J(w)}{z-w}\ ,
\end{align}
where $[t^A,t^B]={f_{ABC}} t^C$ defines a basis of $\su(N)$ and $[t^I,t^J]={f_{IJK}} t^K$ defines a basis of $\su(M)$, while $t^A_{ab}$ and $t^I_{ij}$ are the matrix elements of the fundamental representation
of $\su(N)$ and $\su(M)$, respectively.
We have chosen the bases to be orthonormal, i.e.\ we have that $\tr t^A t^B = \delta_{AB}$ and $\tr t^I t^J = \delta_{IJ}$. 

Then there are the $NM$ Dirac fermions and their conjugates from the $\so(2 N M)_{1}$ factor, which satisfy the standard OPEs
\begin{equation}
\psi^{a i}(z) \bar{\psi}^{b j}(w) \sim  \frac{\delta_{a b} \delta_{ij}}{z-w} \sim \bar \psi^{a i}(z)\psi^{b j}(w)\ .
\end{equation}
Finally, the OPEs between the currents built out of these fermions defined in~\eqref{eq:twosuM} take the form  
\begin{align}\label{eq:sumn_OPE_k}
K^A(z)K^B(w)&\sim \frac{M \delta_{AB} }{(z-w)^2}+\frac{{f_{ABC}}K^C(w)}{z-w}\ ,\\ \notag
K^I(z)K^J(w)&\sim \frac{N \delta_{IJ} }{(z-w)^2}+\frac{{f_{IJK}} K^K(w)}{z-w}\ ,\; K(z)K(w)\sim \frac{NM}{(z-w)^2}\ ,\\ \notag
K^A(z)\psi^{ai}(w)&\sim \frac{t^A_{ba}\psi^{bi}(w)}{z-w}\ ,\; K^I(z)\psi^{ai}(w)\sim \frac{-t^I_{ij}\psi^{aj}(w)}{z-w}\ ,\;K(z)\psi^{ai}(w)\sim \frac{ \psi^{ai}(w)}{z-w}\ ,\\ \notag
K^A(z)\bar \psi^{ai}(w)&\sim \frac{-t^A_{ab}\bar \psi^{bi}(w)}{z-w}\ ,\; K^I(z)\bar \psi^{ai}(w)\sim \frac{t^I_{ji}\bar \psi^{aj}(w)}{z-w}\ ,\;  K(z)\bar \psi^{ai}(w)\sim \frac{-\bar \psi^{ai}(w)}{z-w}\ .
\end{align}


\end{document}